\documentclass[prb,showpacs,preprint,showemail]{revtex4}
\usepackage{amsmath}
\usepackage{graphicx}
\begin{document}
\title{Spin-filtering effect in the transport through a single-molecule magnet Mn$_{12}$ bridged between
metallic electrodes}
\author{Salvador Barraza-Lopez$^{1}$\footnote{Present address: School of Physics, Georgia Institute
of Technology, Atlanta, GA 30332},
Kyungwha Park$^1$,
V\'ictor Garc\'ia-Su\'arez$^{2}$, and Jaime Ferrer$^3$}
\affiliation{$^1$Department of Physics, Virginia Polytechnic Institute and State
University. Blacksburg VA, 24061 \\
$^2$Department of Physics, Lancaster University, Lancaster, LA1 4YB, United Kingdom \\
$^3$Departamento de Fisica, Universidad de Oviedo, 33007 Oviedo, Spain }
\begin{abstract}
Electronic transport through a single-molecule magnet Mn$_{12}$ in a two-terminal
set up is calculated using the non-equilibrium Green's function method in conjunction
with density-functional theory. A single-molecule magnet Mn$_{12}$ is bridged between
Au(111) electrodes via thiol group and alkane chains such that its magnetic easy axis
is normal to the transport direction. A computed spin-polarized transmission coefficient
in zero-bias reveals that resonant tunneling near the Fermi level occurs through
some molecular orbitals of majority spin only. Thus, for low bias voltages, a spin-filtering
effect such as only one spin component contributing to the conductance, is expected.
This effect would persist even with inclusion of additional electron correlations.
\end{abstract}
\date{\today}

\maketitle


Electronic transport through single-molecule magnets (SMMs) has recently been measured
in three-terminal setups or using scanning tunneling microscope (STM).
\cite{HEER06,JO06,HEND07,VOSS07} Both of the experimental techniques face their own
challenges in transport measurements, so do theoretical and computational studies of the
systems. So far, there were no experimental or theoretical inputs on the interfaces
between SMMs and electrodes. It is also extremely difficult to control these interfaces.
SMMs can be differentiated from organic molecules in the sense that they have large
magnetic moments and large magnetic anisotropy barriers caused by spin-orbit coupling.
In addition, SMMs differ from magnetic clusters comprising magnetic elements only
because the magnetic ions in SMMs are interacting with each other via organic/inorganic
ligands. It was shown that the magnetic properties of SMMs change significantly with
the number of extra electrons added to the molecules,\cite{PARK04-3,EPPL95}
which may greatly impact their transport properties. Several theoretical studies
\cite{GHKIM04,ROME06-3,ELST06} on the transport through SMMs, have been, so far,
based on many-body model Hamiltonians considering an important role of strong
correlations in SMMs. However, in these model Hamiltonian studies, effects of
interfaces and molecular geometries that also play a crucial role in transport
were not properly included. In this sense, first-principles calculations could
complement the existing many-body Hamiltonian studies.

\vspace{0.1truecm}
\begin{figure}[h]
\includegraphics[width=8.5 cm, height=3.5 cm]{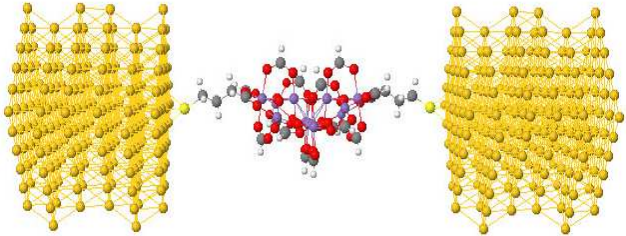}
\caption{(Color online). Extended molecule or scattering region consisting
of SMM Mn$_{12}$ (center) attached to Au layers via S atoms and
alkane chains. Semi-infinite Au electrodes are considered in
the calculations (not shown).}
\label{fig:one}
\end{figure}

In this work, we present first-principles calculations of transport properties
through a SMM Mn$_{12}$ when the molecule is bridged between Au(111) electrodes.
We identify plausible pathways of the transport within the molecule and find a spin-filtering
effect in the transport caused by the nature of Mn$_{12}$ molecular orbitals responsible
for resonant tunneling. For this study, we use the non-equilibrium Green's function method
and the density-functional theory (DFT), with the {\tt SIESTA}-based \cite{SIESTA}
quantum transport code, {\tt SMEAGOL}.\cite{SMEAGOL,FERN06}
To include additional electron correlations, we also add a Hubbard-like $U$ term
to our DFT calculations using {\tt VASP}\cite{VASP}. Our main results reported in this work do not change with
the addition of $U$ term except for an increased HOMO-LUMO gap comparable to the
experimental value.\cite{VOSS07}
In our study the following assumptions are made: (1) Despite intermediate transient
states, the system eventually reaches the stable state. (2) Currents can be obtained from
the methodology based on the ground-state DFT even when a bias voltage is applied.
(3) Interactions of molecules with heat baths are not included. As discussed in
Ref.\onlinecite{SMEAGOL}, we divide the whole system into two parts: (i) bulk left and right
Au electrodes, and (ii) a scattering region consisting of several Au layers, a SMM Mn$_{12}$,
and two linker molecules (S atoms and alkane chains).
The scattering region [also called extended molecule (EM)] is shown in Fig.\ref{fig:one}.
The bulk electrodes are treated semi-infinite and their electronic structures are computed
using {\tt SIESTA}. A current is expressed as \cite{MEIR92}
$I=(e/h) \int dE \: \: \:  T(E,V) (f(E-\mu_L) - f(E-\mu_R))$, where
$T(E,V)$ is a transmission coefficient, $V$ is a bias voltage, and $f(E-\mu_L)$ and
$f(E-\mu_R)$ are the Fermi functions for the left and right electrodes with chemical
potentials $\mu_L$ and $\mu_R$, respectively. $T(E,V)$ can be cast as
\begin{eqnarray}
T(E,V)= \Gamma_L(E,V) G_M^{\dag}(E,V) \Gamma_R(E,V) G_M(E,V)
\end{eqnarray}
where $\Gamma_{L,R}$ is the density of states of the left or right electrode
and $G_M$ is the Green's function of the EM. Within the non-equilibrium Green's function
method, $G_M$ is solved self-consistently in the context of the DFT.

For spin-polarized DFT calculations with {\tt SIESTA} we use Perdew-Burke-Ernzerhof (PBE)
generalized-gradient approximation (GGA) \cite{PERD96} for exchange-correlation potential.
We generate Troullier-Martins pseudopotentials \cite{TROU91} for Au, Mn, S, O, C, and H with scalar
relativistic terms and core corrections except for H, as well as basis sets for each element.
For Mn atoms 3p orbitals are included in the valence states. 
For an isolated Mn$_{12}$
the molecular orbitals near the Fermi level obtained using the generated basis sets and pseudopotentials
agree well within 0.1~eV with those using other DFT codes such as {\tt NRLMOL} \cite{NRLMOL} and
{\tt VASP}. We also compute the magnetic anisotropy barrier for Mn$_{12}$ by including
the spin-orbit interaction self-consistently in DFT calculations with a version of {\tt SIESTA} that
includes spin-orbit coupling \cite{FERN06}. The barrier we obtain agrees with the experimental value
\cite{BARR97}.
All of these tests reveal that our
generated pseudopotentials and basis sets are good enough to study the system of interest.

\vspace{0.1truecm}
\begin{figure}
\includegraphics[width=8.5cm, height=6.0cm]{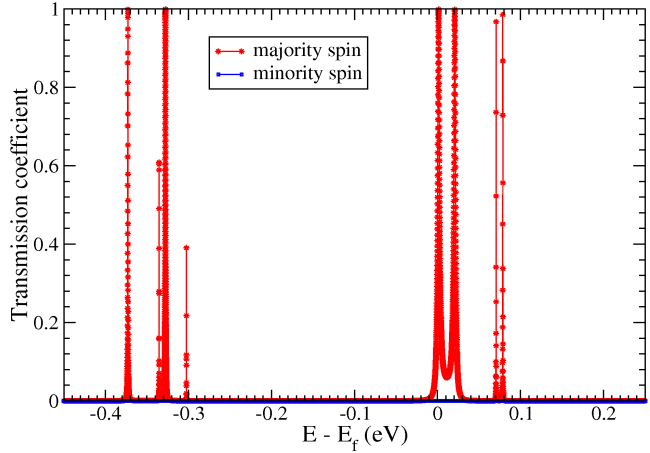}
\caption{(Color online). Spin-polarized transmission coefficient in zero bias.}
\label{fig:three}
\end{figure}

To construct the bulk leads, we use an optimum bulk lattice constant of 4.166~\AA,~resulting in a vertical distance
between adjacent layers of 2.405~\AA.~The electronic structure and self-energies of Au(111) electrodes are computed
using {\tt SIESTA}. For the EM we consider a geometry in which
a Mn$_{12}$ molecule is oriented such that its magnetic easy axis is normal to the transport direction ($z$ axis) and in which
six Au layers are included on each side of the Mn$_{12}$ molecule (Fig.~\ref{fig:one}). This geometry completely
covers a Mn$_{12}$ molecule and prevents different Mn$_{12}$ molecules from interacting with each other.
Mn$_{12}$ molecules are not directly chemically bonded to Au, and so thiol groups and alkane chains are
used to attach Mn$_{12}$ to Au (Fig.~\ref{fig:one}).
In the EM, $3\times 3 \times 1$ $k$ points are sampled. The periodic boundary conditions are imposed on the EM along
all directions to self-consistently solve for the Green's function of the EM.

\vspace{0.1truecm}
\begin{figure}
\includegraphics[width=8.5cm, height=6.3cm]{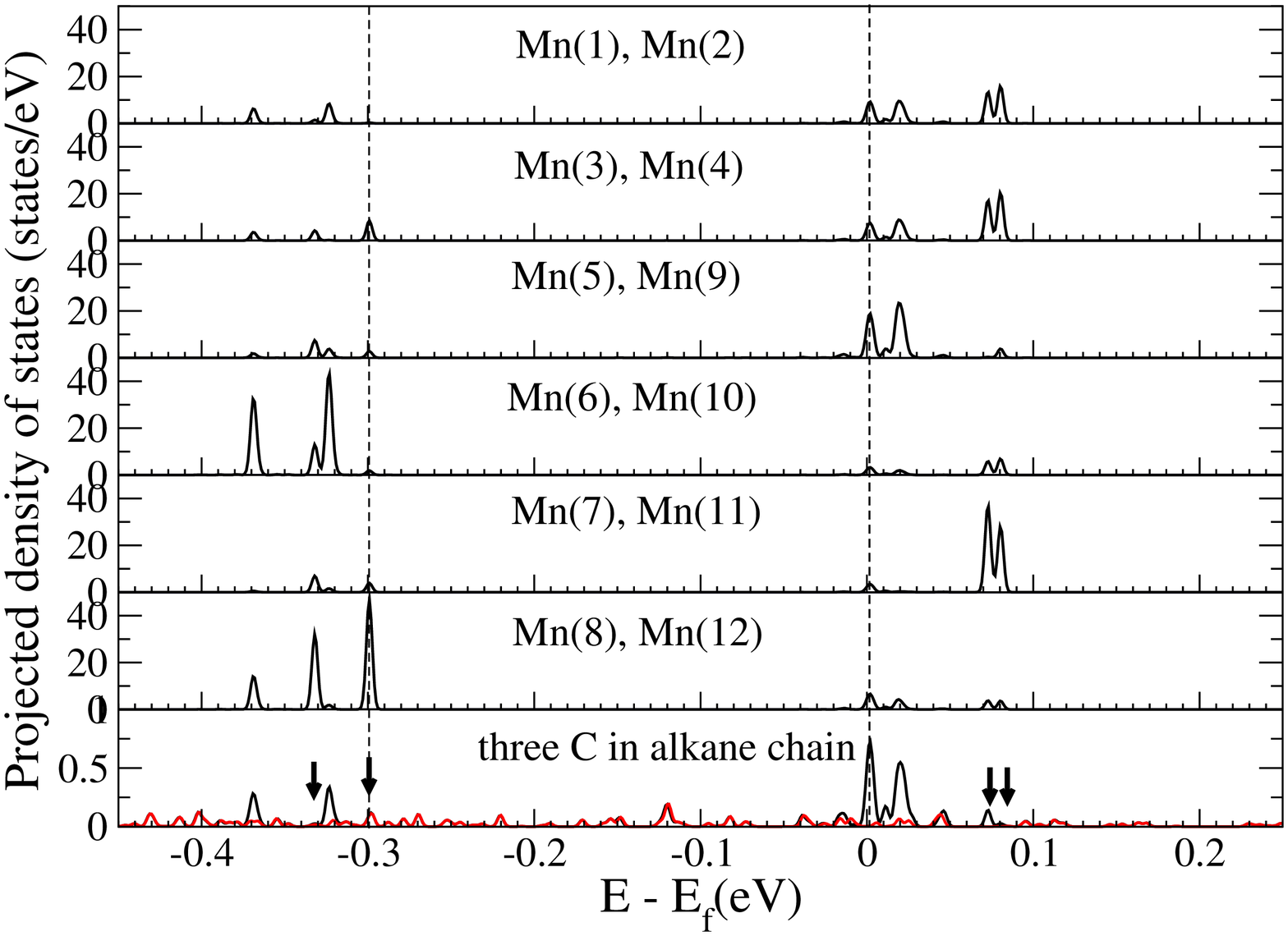}
\caption{(Color online). Spin-polarized density of states of the EM projected onto Mn $d$ and C $p$ orbitals
(majority spin: black, minority spin: red). Minority-spin densities for Mn are zero in the
energy range shown. The Mn ions are labeled in Fig.~\ref{fig:positions}. The dashed lines
(arrows) are for the HOMO and LUMO (HOMO-2, HOMO, LUMO+2, LUMO+3) of Mn$_{12}$
from the left. The LUMO is slightly above the Fermi level.}
\label{fig:PDOS}
\end{figure}

\vspace{0.1truecm}

\begin{figure}
\includegraphics[width=6.0cm, height=3.0cm]{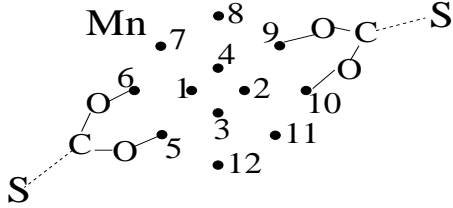}
\caption{Positions of Mn ions (filled circles),  few O, C, and S atoms. The dashed lines represent
the alkane chains. The S atoms are directly bonded to the Au leads.}
\label{fig:positions}
\end{figure}

The computed spin-polarized transmission coefficient in zero bias is shown as a function of energy relative
to the Fermi level $E_f$ in Fig.~\ref{fig:three}. The peaks in the transmission coefficient are very narrow
due to weak coupling between the electrodes and Mn$_{12}$, as discussed in our previous DFT calculation.\cite{SALV07}
Thus, one can identify a one-to-one mapping between individual transmission peaks
and molecular orbitals. The transmission coefficient from minority-spin electrons is zero in the energy region
shown in Fig.~\ref{fig:three}. The first minority-spin transmission peak appears
at 0.71~eV above $E_f$. So contributions to the transmission near $E_f$ are from majority-spin electrons only. This is
because only majority-spin molecular orbitals are near $E_f$ and they are well separated from minority-spin
molecular orbitals. This feature may be seen in the spin-polarized projected density of states onto Mn $d$
and C $p$ orbitals (Figs.~\ref{fig:PDOS} and \ref{fig:positions}). The minority-spin densities of Mn $d$ orbitals
are zero in the energy range shown in Fig.~\ref{fig:PDOS}. Compared to fourfold symmetric Mn $d$ orbitals
in an isolated Mn$_{12}$, the Mn $d$ orbitals in the EM reveal twofold symmetry because of broken symmetry caused
by the leads (Fig.~\ref{fig:PDOS}). By comparing Fig.~\ref{fig:PDOS} to Fig.~\ref{fig:three}, we find that the
four transmission peaks between $E$=0 and 0.1~eV are from the majority-spin lowest-unoccupied-molecular-orbital
(LUMO) and the three levels right above the LUMO (LUMO+n, n=1,2,3), while the four peaks between $E$=-0.4 and
-0.3~eV are from the majority-spin highest-occupied-molecular-orbital (HOMO) and the three levels right below
the HOMO (HOMO-n, n=1,2,3). The HOMO is mainly from Mn(8) and Mn(12) and the HOMO-1 and HOMO-3 are from Mn(6)
and Mn(10). The LUMO and LUMO+1 are from Mn(5) and Mn(9) and the LUMO+2 and LUMO+3 are from Mn(7) and Mn(11).
Although a corresponding transmission peak appears at the energy level of each molecular orbital
in the range, not all of the orbitals would significantly contribute to the current because of the extremely narrow widths
of peaks. The peaks at $E$=-0.335, -0.35, 0.07, and 0.08~eV, have widths of less than 10$^{-5}$~eV, and so
they would not contribute much to the current. The widths of transmission peaks depend on the extent of the
electron density overlap or level broadening along the transport pathways. For example, for electrons to be
transmitted through Mn$_{12}$, they must tunnel through the alkane chains. The spin-polarized density of states
projected onto the three C atoms in each alkane chain shows small peaks at the energies of the HOMO-2, HOMO,
LUMO+2, LUMO+3 as indicated by the arrows in the bottom panel of Fig.~\ref{fig:PDOS}, resulting in extremely
narrow transmission peaks. Thus, more electrons would tunnel through the LUMO, LUMO+1, HOMO-1, and HOMO-3.

For comparison with experimental data, strong correlation effects
which are lacking in GGA, should be included. Insight on these effects on transport is provided
from GGA+U studies on an isolated Mn$_{12}$. Our previous GGA+U study on a neutral Mn$_{12}$ \cite{SALV08}
suggests that the HOMO-LUMO gap shown in Figs.~\ref{fig:three} and \ref{fig:PDOS} would increase
so that a higher bias voltage would be needed to charge the Mn$_{12}$ molecule than appeared
in Fig.~\ref{fig:three}. However, with a downward shift of occupied levels and upward shift of
unoccupied levels, the main features including spin filtering would persist. When the Mn$_{12}$ molecule 
is singly charged with inclusion of $U$, we find that the level corresponding to the LUMO of the 
neutral Mn$_{12}$ is now filled so that the energy gap for the charged Mn$_{12}$ becomes very small. 
Thus, what we learned from the zero-bias case would be still relevant for non-zero bias cases to large
extent. Further studies on the charged Mn$_{12}$ molecule in transport are in progress.

In summary, we have investigated transport properties through a Mn$_{12}$ molecule bridged
between Au(111) electrodes using the non-equilibrium Green's function method and spin-polarized
DFT. We found that a spin-filtering effect occurs in the transmission for non-ferromagnetic
electrodes. In addition, not all of the molecular orbitals near $E_f$ involved with the resonant
tunneling, equally contribute to the current.



K.P. was supported by the Jeffress Memorial Trust Funds and NSF DMR-0804665.
J.F. was supported by MEC FIS2006-12117.
Computational support was provided by the SGI Altix Linux Supercluster
at the National Center for Supercomputing Applications under DMR060009N
and by Virginia Tech Linux clusters and Advanced Research Computing.


\begin{thebibliography}{10}

\bibitem{HEER06}
H. B. Heersche, Z. de Groot, J. A. Folk, H. S. van der Zant, C. Romeike, M. R.
Wegewijs, L. Zobbi, D. Barreca, E. Tondello, and A. Cornia,
Phys. Rev. Lett. {\bf 96}, 206801 (2006)

\bibitem{JO06}
M.-H.\ Jo, J.~E.\ Grose, K.\ Baheti, M.~M.\ Deshmukh, J.~J.\ Sokol,
E.~M.\ Rumberger, D.~N.\ Hendrickson, J.~R.\ Long, H.\ Park, and
D.~C.\ Ralph, Nano Lett. {\bf 6}, 2014 (2006).

\bibitem{HEND07}
J.J. Henderson, C.M. Ramsey, E. del Barco, A. Mishra, and G. Christou,
J. Appl. Phys. {\bf 101}, 09E102 (2007).

\bibitem{VOSS07}
S. Voss, M. Fonin, U. Rudiger, M. Burgert, and U. Groth,
Appl. Phys. Lett. {\bf 90}, 133104 (2007).

\bibitem{PARK04-3}
K. Park and M. R. Pederson, Phys. Rev. B {\bf 70}, 054414 (2004).

\bibitem{EPPL95}
H.~J.\ Eppley {\it et al.},
J.\ Am.\ Chem.\ Soc.\ {\bf 117}, 301 (1995);
S.M.J. Aubin {\it et al.},
Inorg. Chem. {\bf 38}, 5329 (1999).

\bibitem{GHKIM04}
G.-H.\ Kim and T.-S.\ Kim, Phys. Rev. Lett. {\bf 92}, 137203 (2004).

\bibitem{ROME06-3}
C. Romeike, M. R. Wegewijs, and H. Schoeller, Phys. Rev. Lett. {\bf 96}, 196805 (2006).

\bibitem{ELST06}
F.\ Elste and C.\ Timm, Phys. Rev. B {\bf 73}, 235305 (2006).


\bibitem{SIESTA}
J.~M.\ Soler {\it et al.},
J. Phys.: Condens. Matter {\bf 14}, 2745 (2002);
J. Junquera {\it et al.},
Phys. Rev. B {\bf 64}, 235111 (2001);
P. Ordej\'on {\it et al.},
Phys. Rev. B {\bf 51}, 1456 (1995).

\bibitem{SMEAGOL}
A.R. Rocha, V.M. Garc\'ia-Su\'arez, S. Bailey, C. Lambert, J. Ferrer,
and S. Sanvito, Phys. Rev. B {\bf 73}, 085414 (2006).

\bibitem{FERN06}
L. Fern\'andez-Seivane, M. A. Oliveira, S. Sanvito, and J. Ferrer,
J. Phys.: Condens. Matter {\bf 18}, 7999 (2006).

\bibitem{VASP}
G. Kresse and J. Furthm\"uller, Phys. Rev. B {\bf 54}, 11169 (1996);
G. Kresse and J. Furthm\"uller, Comp. Mat. Sci. {\bf 6}, 15 (1996).

\bibitem{MEIR92}
Y. Meir and N. S. Wingreen, Phys. Rev. Lett. {\bf 68}, 2512 (1992).

\bibitem{PERD96}
J.~P.\ Perdew, K.\ Burke, and M.\ Ernzerhof, Phys. Rev. Lett. {\bf 77},
3865 (1996).

\bibitem{TROU91}
N. Troullier and J. L. Martins, Phys. Rev. B {\bf 43}, 1993 (1991).


\bibitem{NRLMOL}
M.~R.\ Pederson and K.~A.\ Jackson, Phys. Rev. B {\bf 41}, 7453 (1990);
K.~A.\ Jackson and M.~R.\ Pederson, {\it ibid.} {\bf 42}, 3276 (1990);
D.~V.\ Porezag, Ph.D. thesis, Chemnitz Technical Institute, 1997.


\bibitem{BARR97}
A.~L.\ Barra, D.\ Gatteschi, and R.\ Sessoli, Phys. Rev. B {\bf 56}, 8192 (1997).

\bibitem{SALV07}
S. Barraza-Lopez, M. C. Avery, and K. Park, Phys. Rev. B {\bf 76}, 224413 (2007).

\bibitem{SALV08}
S. Barraza-Lopez, M. C. Avery, and K. Park, J. Appl. Phys. {\bf 103}, 07B907 (2008).



\end{thebibliography}
\end{document}